%% file: LL14-procs-DY-Oaas_v2.tex
\documentclass{PoS}

\usepackage{cite}
\usepackage{epsfig}

\include{setup}

\title{\boldmath{${\cal O}(\alpha_{\mathrm{s}}\alpha)$} corrections to Drell--Yan processes \\ in the resonance region}

\ShortTitle{{${\cal O}(\alpha_{\mathrm{s}}\alpha)$} corrections to Drell--Yan processes in the resonance region}

\author{\speaker{Stefan Dittmaier}\\
        Albert-Ludwigs-Universit\"at Freiburg, Physikalisches Institut,
        D-79104 Freiburg, Germany\\
        E-mail: \email{stefan.dittmaier@physik.uni-freiburg.de}}

\author{Alexander Huss\\
        Albert-Ludwigs-Universit\"at Freiburg, Physikalisches Institut,
        D-79104 Freiburg, Germany\\
        E-mail: \email{alexander.huss@physik.uni-freiburg.de}}

\author{Christian Schwinn\\
        Albert-Ludwigs-Universit\"at Freiburg, Physikalisches Institut,
        D-79104 Freiburg, Germany\\
        E-mail: \email{christian.schwinn@physik.uni-freiburg.de}}

\abstract{Drell--Yan-like W-boson and Z-boson production in the resonance region
allows for some high-precision measurements that are crucial to carry experimental tests of
the Standard Model to the extremes, such as the determinations of the W-boson mass and
the effective weak mixing angle.
We describe how the Standard Model prediction can be successfully performed in terms
of a consistent expansion about the resonance pole, which classifies the corrections in
terms of factorizable and non-factorizable contributions. The former can be attributed to
the W/Z production and decay subprocesses individually, while the latter link production and decay
by soft-photon exchange. At next-to-leading order we compare the full electroweak
corrections with the pole-expanded approximations, confirming the validity of the 
approximation. At ${\cal O}(\alpha_{\mathrm{s}}\alpha)$, we describe the concept of the
expansion and report on results on the non-factorizable contributions, which
turn out to be phenomenologically negligible. 
Moreover, we present first (preliminary) results on the dominant factorizable 
${\cal O}(\alpha_{\mathrm{s}}\alpha)$ corrections, which originate from the interplay of 
initial-state QCD and final-state electroweak corrections. 
The naive factorization of NLO QCD and NLO EW corrections approximates
the ${\cal O}(\alpha_{\mathrm{s}}\alpha)$ corrections to the W-boson transverse-mass distribution,
but, e.g., not the distribution in the lepton transverse momentum.
}

\FullConference{ Loops and Legs in Quantum Field Theory - LL 2014,\\
		 27 April - 2 May 2014 \\
		 Weimar, Germany }

\begin{document}

\section{Introduction}

Drell--Yan-like W- or Z-boson production is among the most important
standard candle processes at the LHC.
Apart from delivering important information on parton distributions
and allowing for the search for new gauge bosons in the high-mass range,
these processes allow for high-precision measurements in the resonance regions.
The weak mixing angle might be measured with LEP precision,
and the W-boson mass $\MW$ with an accuracy exceeding 
$10\MeV$.

In the past two decades, great effort was made in the theory community
to deliver precise predictions matching the required accuracy
(for a list of references, see \citere{Dittmaier:2014qza}).
QCD corrections are known up to next-to-next-to-leading order,
electroweak (EW) corrections up to next-to-leading order (NLO).
Both on the QCD and on the EW side, there are further refinements such as
leading higher-order effects, resummations, matched parton showers.
In view of fixed-order calculations, the largest missing piece seems to be
the mixed QCD--EW corrections of ${\cal O}(\alpha_{\mathrm{s}}\alpha)$.
Knowing the contribution of this order will also answer the question how
to properly combine QCD and EW corrections in predictions.
In \citere{Balossini:2009sa} this issue is quantitatively discussed with special
emphasis on observables that are relevant for the $\MW$ determination,
revealing percent corrections of ${\cal O}(\alpha_{\mathrm{s}}\alpha)$
that should be calculated.
First steps towards this direction have been taken
by calculating two-loop contributions~\cite{Kotikov:2007vr},
the full ${\cal O}(\alpha_{\mathrm{s}}\alpha)$ correction to the W/Z-decay widths~\cite{Czarnecki:1996ei},
and the full ${\cal O}(\alpha)$ EW corrections to W/Z+jet production
including the W/Z decays~\cite{Denner:2009gj}.

In this short article, we briefly report on our effort~\cite{Dittmaier:2014qza} 
to calculate the ${\cal O}(\alpha_{\mathrm{s}}\alpha)$
corrections to Drell--Yan processes in the resonance region via the 
so-called {\it pole approximation}, which is based on a systematic expansion
about the resonance pole.
Specifically, we sketch the salient features of the approach, discuss its success
at NLO, and describe results on the so-called non-factorizable contributions
at ${\cal O}(\alpha_{\mathrm{s}}\alpha)$, which comprise the most delicate 
contribution to the PA.
Moreover, we present first (preliminary) results on the dominant factorizable
corrections, originating from the interplay of
initial-state QCD and final-state electroweak corrections.

\section{Pole approximation for NLO corrections}

The general idea~\cite{Stuart:1991xk}
of a pole approximation (PA) for any Feynman diagram with a single
resonance is the systematic isolation of all parts that are enhanced by
a resonance factor $1/(p^2-M_\PV^2+\ri\MV\Gamma_\PV)$, where
$p$, $\MV$, and $\Gamma_\PV$ are the momentum, mass, and width 
of the resonating particle~$\PV$, respectively. 
In Drell--Yan production, $\PV$ stands for a W or a Z~boson.
For W~production different variants of PAs have been 
suggested and discussed at NLO already in 
\citeres{Wackeroth:1996hz,Dittmaier:2001ay}.
For the virtual corrections we follow the PA approach of \citere{Dittmaier:2001ay}.
Note, however, that we apply the PA to the real corrections as well, in contrast to
\citere{Dittmaier:2001ay} where they were based on full matrix elements.

Schematically each transition amplitude has the form
\beq
\M=\frac{W(p^2)}{p^2-M_V^2+\Sigma(p^2)}+N(p^2),
\eeq
with functions $W$ and $N$ describing resonant and non-resonant parts, respectively,
and $\Sigma$ denoting the self-energy of $V$.
The resonance of $\M$ is isolated in a gauge-invariant way as follows,
\beq
\M =
\frac{W(\mu_V^2)}{p^2-\mu_V^2}\,\frac{1}{1+\Si'(\mu_V^2)}
+\left[\frac{W(p^2)}{p^2-M_V^2+\Si(p^2)}
- \frac{W(\mu_V^2)}{p^2-\mu_V^2}\,\frac{1}{1+\Si'(\mu_V^2)} \right] + N(p^2),
\label{eq:PA}
\eeq
where $\mu_V^2=M_V^2-\ri M_V\Ga_V$ is the gauge-invariant location of the
propagator pole in the complex $p^2$ plane.
Equation~\refeq{eq:PA} can serve as a basis for the gauge-invariant introduction of 
the finite decay width in the resonance propagator, thereby defining the so-called
{\it pole scheme}. In this scheme the term in square brackets is 
perturbatively expanded in the coupling $\alpha$ including terms up to ${\cal O}(\alpha)$,
while the full $p^2$ dependence is kept.
An application of this scheme to Z-boson production is, e.g.,
described in \citere{Dittmaier:2009cr} in detail.

\begin{figure}
\epsfig{file=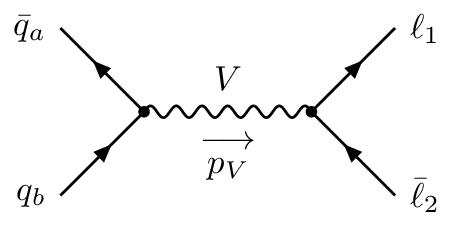,scale=0.75}
\hfill
\epsfig{file=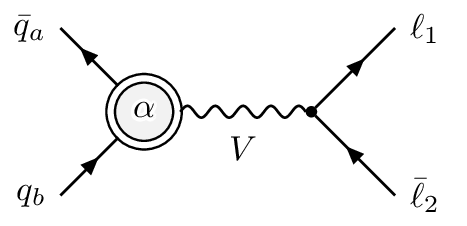,scale=0.75}
\hfill
\epsfig{file=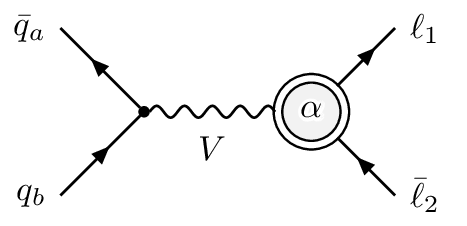,scale=0.75}
\hfill
\epsfig{file=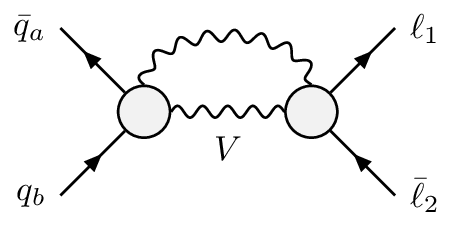,scale=0.75}
\\[-.5em]
\hspace*{3.5em} (a) \hfill (b) \hfill (c) \hfill (d) \hspace*{4em}
\caption{Generic diagrams for the lowest-order amplitude (a),
for the EW virtual NLO factorizable corrections to production (b) and decay (c),
as well as for virtual non-factorizable corrections (d), where the empty blobs stand for all relevant
tree structures and the ones with ``$\alpha$'' inside 
for one-loop corrections of ${\cal O}(\alpha)$.}
\label{fig:vNLOgraphs}
\end{figure}

The PA for the amplitude
results from the r.h.s.\ of \refeq{eq:PA} upon neglecting the last, non-resonant term and
asymptotically expanding the term in square brackets 
in $p^2$ about the point $p^2=\mu_\PV^2$, 
where only the leading, resonant term of the expansion is kept.
The first term on the r.h.s.\ of \refeq{eq:PA} defines the so-called
{\it factorizable} corrections in which on-shell production and decay amplitudes
for $V$ are linked by the off-shell propagator; these contributions
are illustrated by diagrams (b) and (c) of \reffi{fig:vNLOgraphs}.
The term on the r.h.s.\ of \refeq{eq:PA} in square brackets contains the so-called
{\it non-factorizable} corrections which receives resonant contributions from all diagrams
where the limit $p^2\to\mu_V^2$ in $W(p^2)$ or $\Si'(p^2)$ would lead to (infrared)
singularities. At NLO, this happens if a soft photon of energy $E_\gamma\lsim\Ga_\PV$
is exchanged between the production process, the decay part, and the intermediate $V$ bosons;
a generic loop diagram is shown in \reffi{fig:vNLOgraphs}(d).
Although in principle the real-emission corrections can be based on full amplitudes
without further approximations,
the consistent application of the PA to both virtual and real corrections
is necessary to make a separate discussion of factorizable
and non-factorizable contributions possible.
\looseness -1

Conceptually the evaluation of the non-factorizable corrections is the
most delicate among all PA contributions. They possess rather interesting features.
When the invariant mass of the resonance is integrated over, their contribution
vanishes~\cite{Fadin:1993dz}, 
i.e.\ they only tend to distort the resonance without changing the normalization
of the cross section.
Since they only involve soft photons, they take the form of a global correction
factor to the lowest-order matrix element squared, with a non-trivial dependence
on the virtuality $(p^2-\mu_\PV^2)$ which gave the corrections their name.
The virtual correction factor can be explicitly calculated with quite
compact results, even for double resonances for which their calculation is described
in detail in the literature~\cite{Melnikov:1995fx}.
The real corrections are better evaluated numerically to keep some flexibility
in the treatment of photons in the event selection, using extended
eikonal currents~\cite{Melnikov:1995fx} 
that take into account the resonance distortion by soft photons
of energy $E_\gamma\lsim\Ga_\PV$.
Collinear singularities do not occur at all, since all relevant diagrams
are of interference type.

In spite of the simple general idea of the PA, its consistent implementation
in higher-order calculations involves subtle details. 
For instance, care has to be taken that subamplitudes appearing before or after 
the $V$ resonance are based on subamplitudes with on-shell $V$ bosons, 
otherwise gauge invariance cannot be guaranteed. 
Setting $p^2=\MV^2$, instead of the problematic complex value $p^2=\mu_\PV^2$,
in the ${\cal O}(\alpha)$ corrections is certainly allowed in ${\cal O}(\alpha)$ approximation.
However, the procedure is not unique, because the phase space is parametrized by
more than one variable. The {\it on-shell projection} $p^2\to\MV^2$ has to be defined
carefully. Different variants may lead to results that differ within the
intrinsic uncertainty of the PA, which is of ${\cal O}(\alpha/\pi\times\Gamma_\PV/\MV)$
in the resonance region when applied to ${\cal O}(\alpha)$ corrections.
However, care has to be taken that virtual and real corrections still match properly
in the (soft and collinear) infrared limits in order to guarantee the cancellation
of the corresponding singularities.

\begin{figure}
\epsfig{file=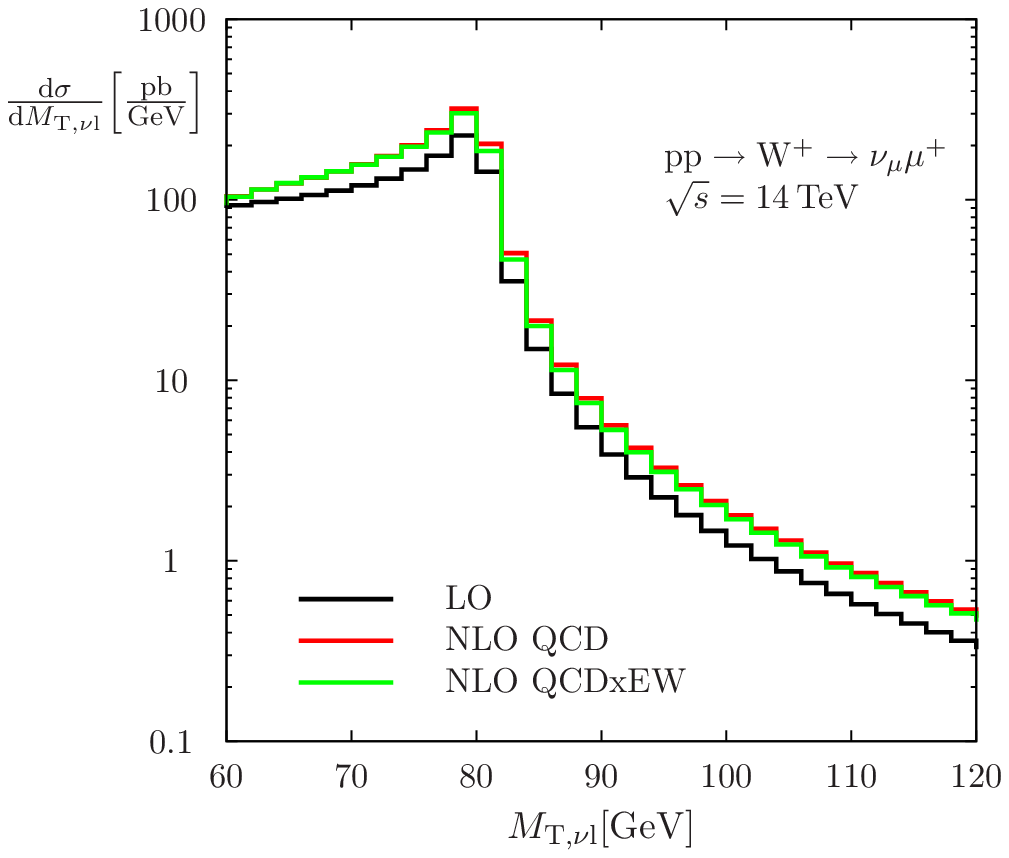,scale=0.7} \hfill \epsfig{file=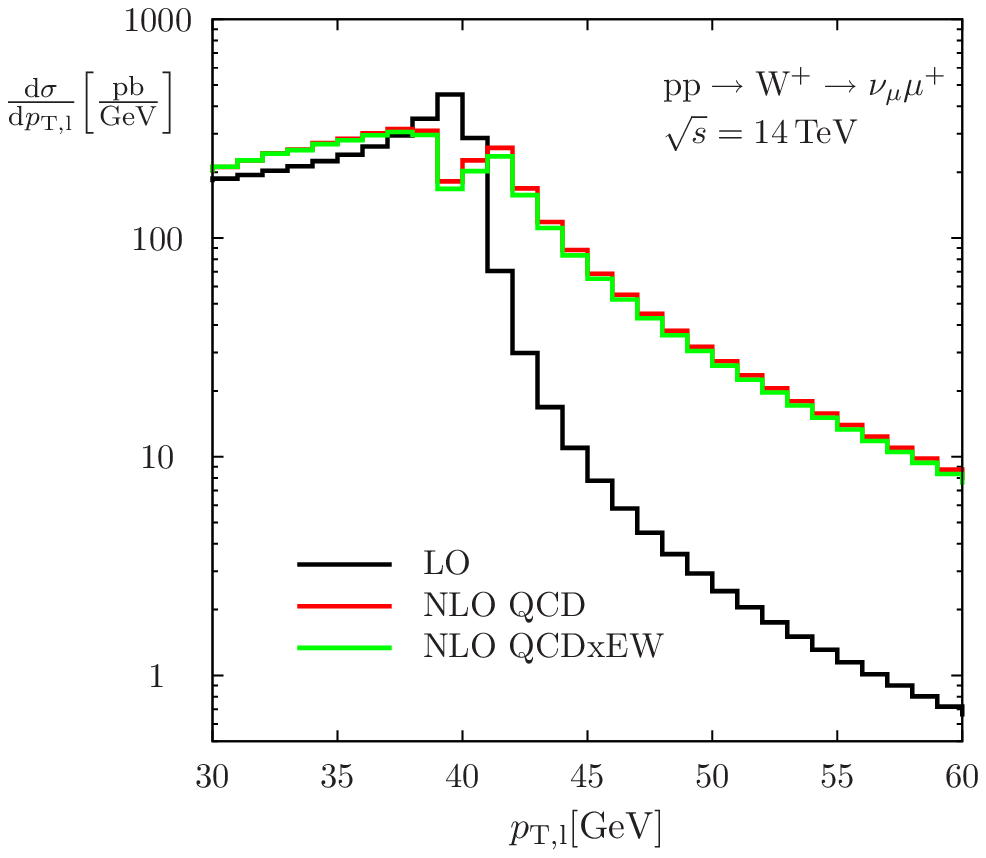,scale=0.7} \hfill
\\[.5em]
\hspace*{1.5em}\epsfig{file=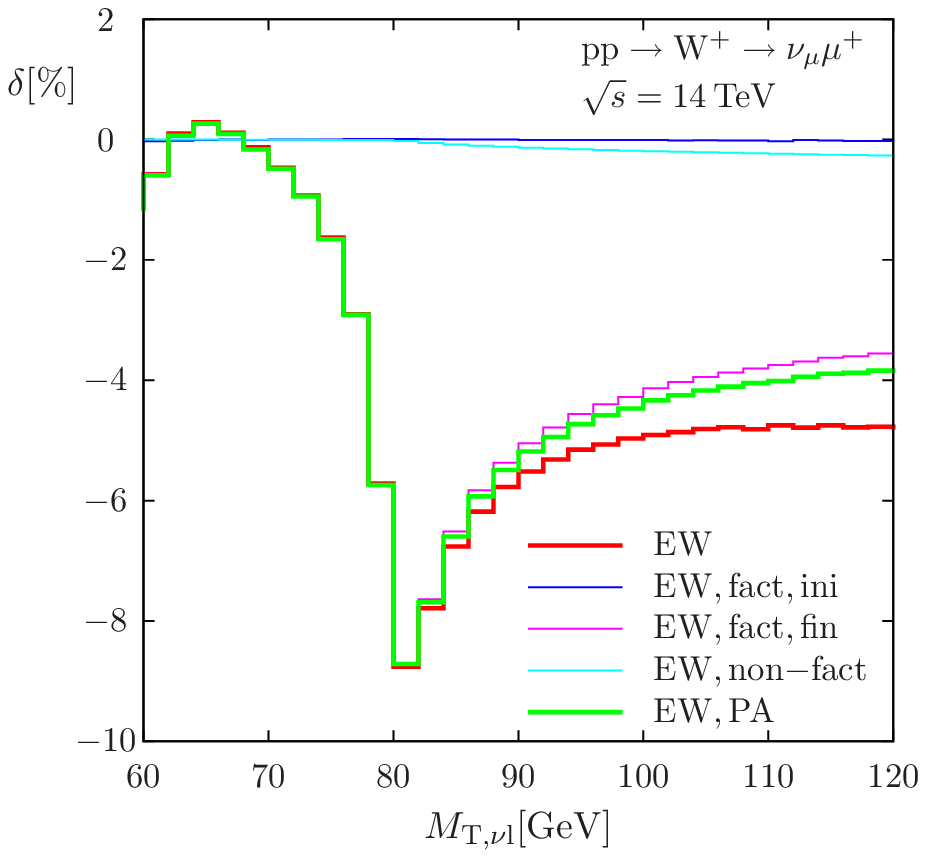,scale=0.7} \hfill \epsfig{file=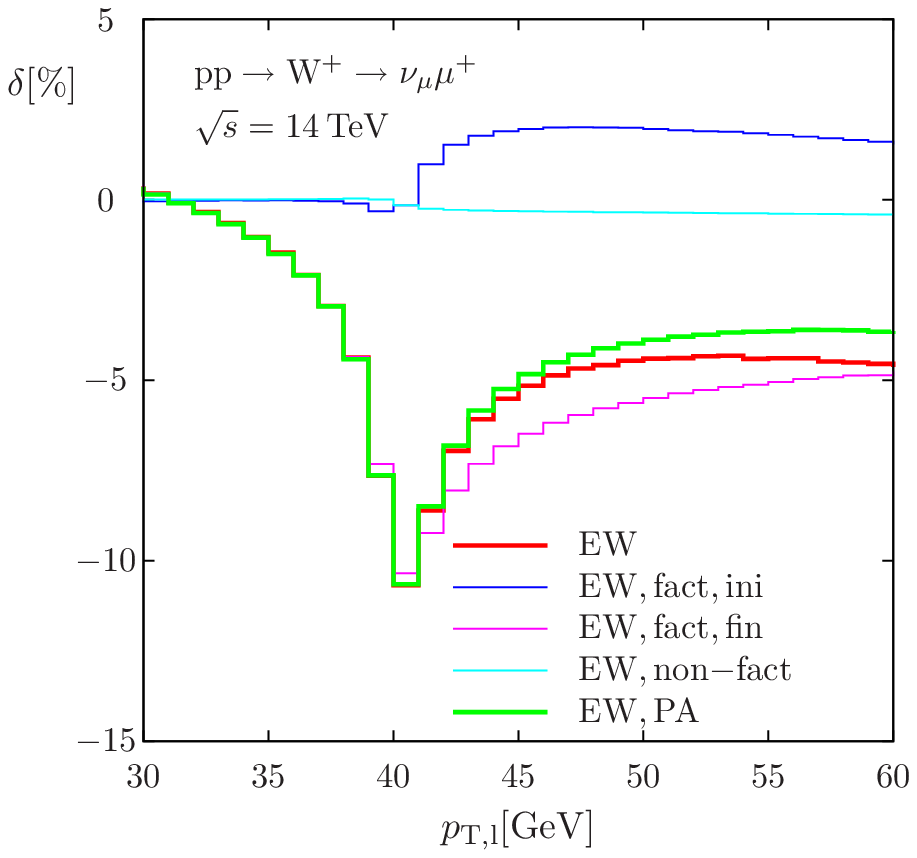,scale=0.7} \hfill
\caption{Distributions in the transverse mass (left) and transverse lepton momentum (right)
for $\PWp$ production at the LHC, with the upper plots showing the absolute distributions
and the lower plots the relative NLO 
EW corrections in PA broken up into its factorizable and non-factorizable parts
(taken from \citere{Dittmaier:2014qza}).}
\label{fig:PA-NLO}
\end{figure}

Based on our results derived in \citere{Dittmaier:2014qza},
Figure~\ref{fig:PA-NLO} exemplarily shows the NLO QCD and EW corrections to the
transverse-mass and transverse-lepton-momentum distributions for $\PWp$ production
at the LHC and, in particular, illustrates the structure and quality of the PA
applied to the EW corrections.
The distributions show the well-known Jacobian peaks at $M_{\rT,\nu\Pl}\sim\MW$ and
$p_{\rT,\Pl}\sim\MW/2$, respectively, which play a central role in the measurement of
the W-boson mass $\MW$ at hadron colliders.
The EW corrections significantly distort the distributions and shift the peak position.
Note also the extremely large QCD corrections above the peak in the $p_{\rT,\Pl}$
distribution, which are induced by the recoil of the W~boson against the hard jet
of the real QCD correction.
The lower panels of \reffi{fig:PA-NLO} compare the full NLO EW corrections
(without photon-induced processes from $q\gamma$ collisions)
to the result of the PA, which is also broken up into factorizable corrections to
the initial/final state and non-factorizable contributions.
Near the Jacobian peaks, the PA turns out to be good within some $0.1\%$.
Interestingly, the impact of the non-factorizable corrections is suppressed to the
$0.1\%$ level and, thus, phenomenologically negligible. A similar conclusion
even holds for the factorizable initial-state corrections when one takes into account that
the percentage correction to the $p_{\rT,\Pl}$ distribution actually should be normalized
to the full cross section including QCD corrections, which are overwhelming above
the Jacobian peak.
Thus, the relevant part of the NLO EW corrections near the peaks entirely results
from the factorizable final-state corrections, where the bulk originates from
collinear final-state radiation from the decay leptons.

More details and results on the PA at NLO
are discussed in \citere{Dittmaier:2014qza}, in particular for $\PZ$~production, for which
the PA works similarly well.

\section{Pole approximation at \boldmath{${\cal O}(\alpha_{\mathrm{s}}\alpha)$}} 

\begin{figure}
\centerline{\epsfig{file=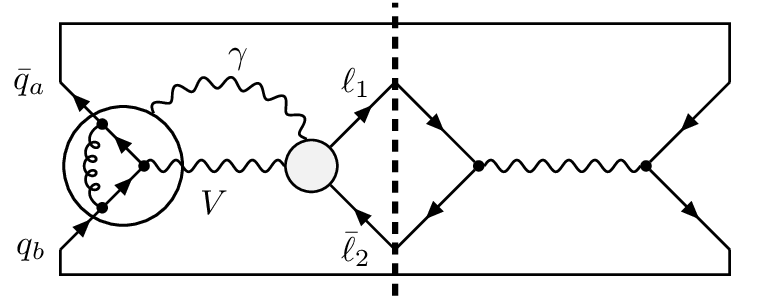,scale=0.75}
\hspace*{3em}
\epsfig{file=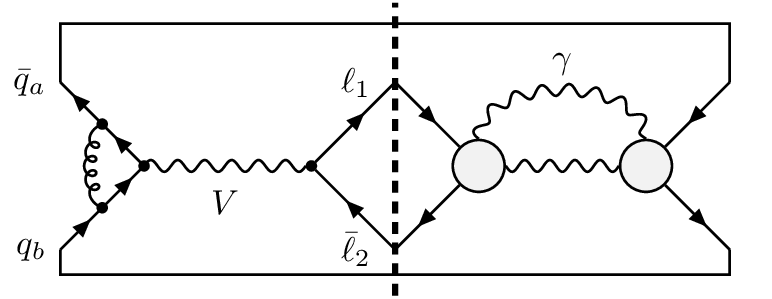,scale=0.75}}
\centerline{\epsfig{file=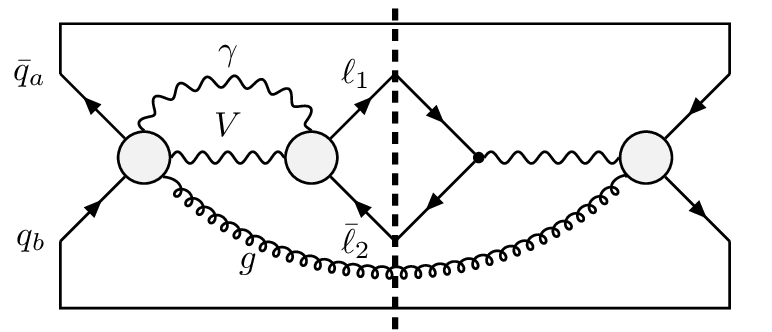,scale=0.75}
\hspace*{3em}
\epsfig{file=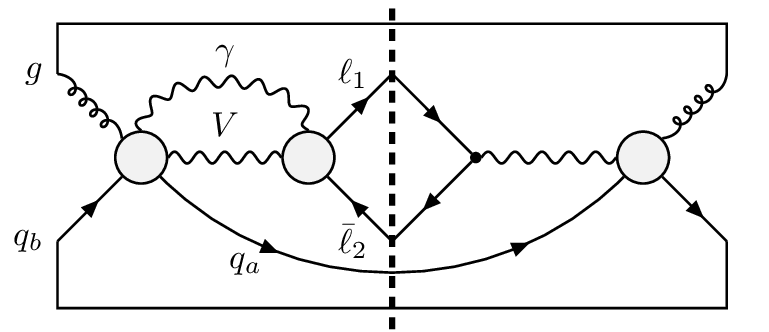,scale=0.75}}
\centerline{\epsfig{file=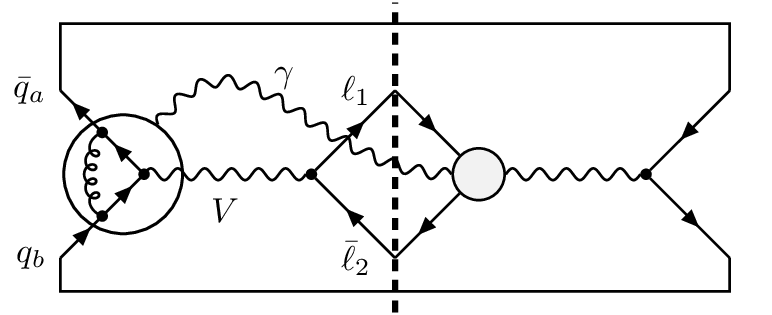,scale=0.75}
\hspace*{3em}
\epsfig{file=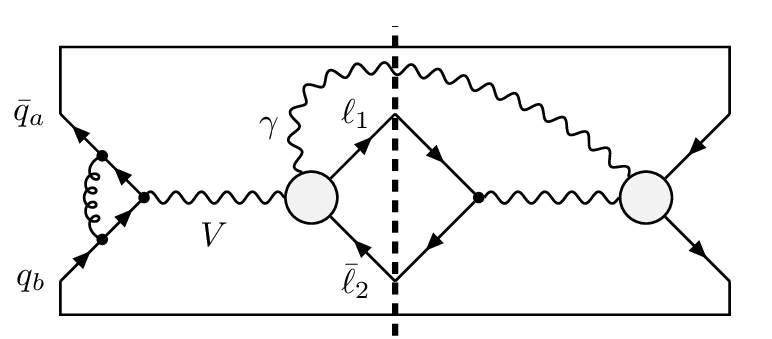,scale=0.75}}
\centerline{\epsfig{file=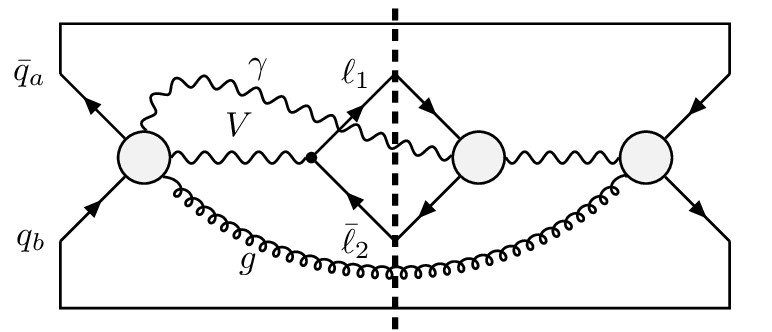,scale=0.75}
\hspace*{3em}
\epsfig{file=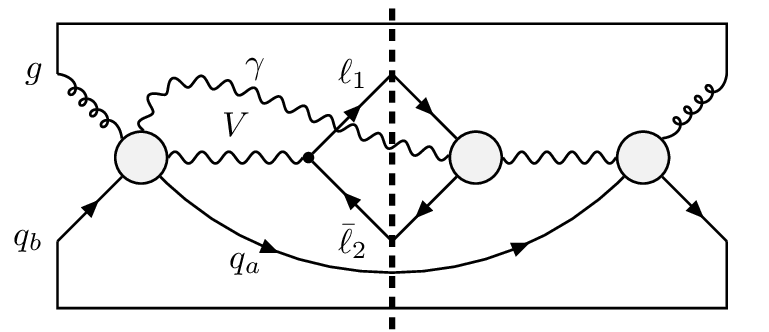,scale=0.75}}
\vspace*{-.5em}
\caption{Generic diagrams for the non-factorizable
corrections of ${\cal O}(\alpha_{\mathrm{s}}\alpha)$.}
\label{fig:nfOaasgraphs}
\end{figure}

\begin{figure}
\centerline{\epsfig{file=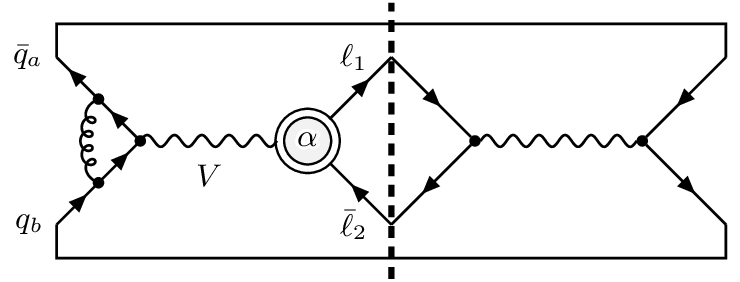,scale=0.75}
\hspace*{3em}
\epsfig{file=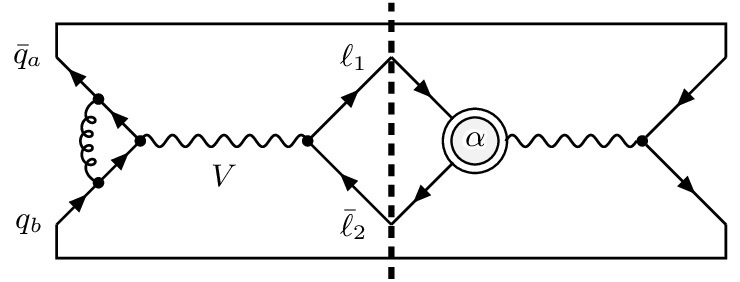,scale=0.75}}
\centerline{\epsfig{file=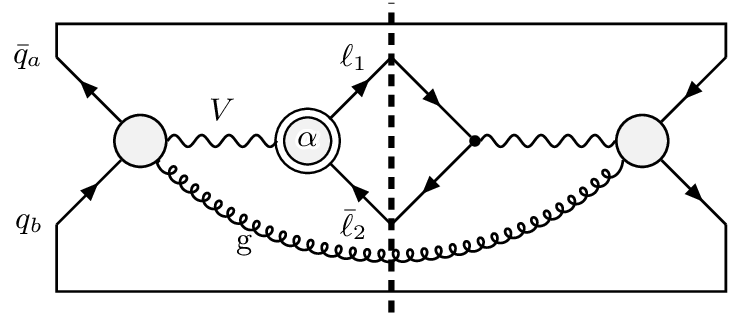,scale=0.75}
\hspace*{3em}
\epsfig{file=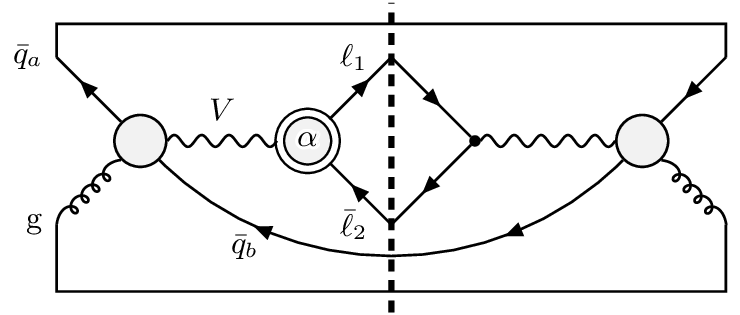,scale=0.75}}
\centerline{\epsfig{file=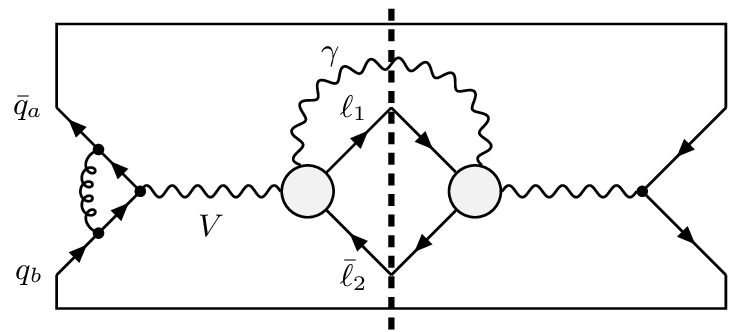,scale=0.75}}
\centerline{\epsfig{file=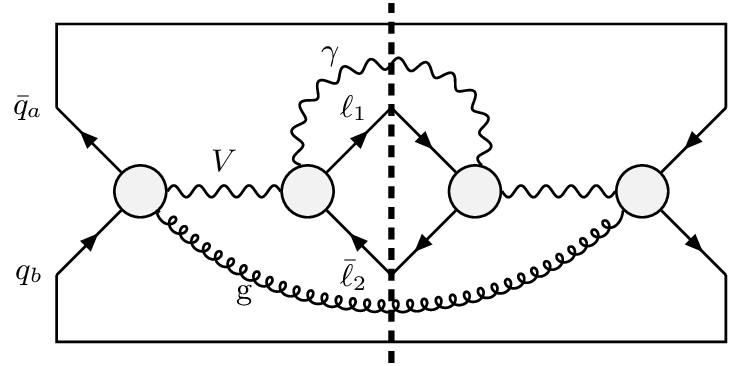,scale=0.75}
\hspace*{3em}
\epsfig{file=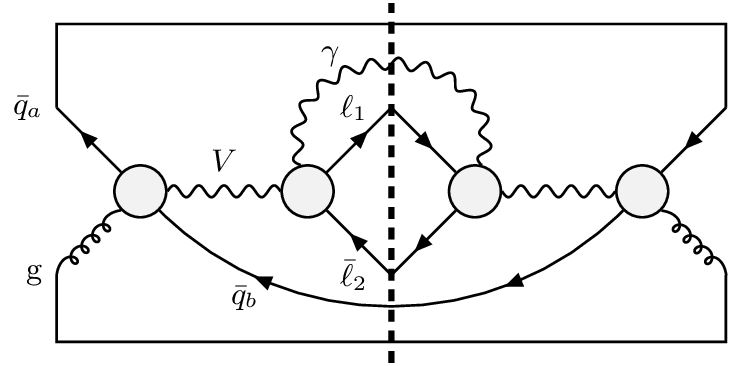,scale=0.75}}
\vspace*{-.5em}
\caption{Generic diagrams for the ${\cal O}(\alpha_{\mathrm{s}}\alpha)$ factorizable
corrections of the ``initial--final'' type.}
\label{fig:IFfactOaasgraphs}
\end{figure}

Though technically more complicated,
the concept of the PA, described at NLO in the previous section, can be carried over
to higher orders in a straightforward way.
The corresponding virtual, real, and mixed virtual--real
contributions involve various different interference diagrams.
Exemplarily we depict the generic graphs for the non-factorizable 
${\cal O}(\alpha_{\mathrm{s}}\alpha)$ corrections in \reffi{fig:nfOaasgraphs}
and for the factorizable contributions combining initial-state QCD and final-state EW
corrections in \reffi{fig:IFfactOaasgraphs}.
\looseness-1


As at NLO, the non-factorizable corrections originate from the virtual exchange or 
real emission of soft photons with energies $E_\gamma\lsim\Gamma_\PV$, however, without
any restriction on the kinematics of virtual gluons or real jet radiation.
In \citere{Dittmaier:2014qza} we have discussed in detail the factorization properties of the virtual and real
photonic parts
of the non-factorizable ${\cal O}(\alpha_{\mathrm{s}}\alpha)$ corrections, which result
from the soft nature of the effect. 
Using gauge-invariance arguments borrowed from the classic YFS paper~\cite{Yennie:1961ad},
we show that this factorization of the photonic factors even hold to any order in the
strong coupling $\alpha_{\mathrm{s}}$.
We have verified this statement diagrammatically and, for the
purely virtual corrections, also with effective-field-theory 
techniques~\cite{Beneke:2003xh}.
Both the virtual and real photonic corrections can be written as correction factors to
squared matrix elements containing gluon loops or external gluons, i.e.\ the necessary 
building blocks are obtained from tree-level and one-loop calculations.
Our numerical study~\cite{Dittmaier:2014qza} shows that the non-factorizable corrections
of ${\cal O}(\alpha_{\mathrm{s}}\alpha)$ below the $0.1\%$ level and, thus, phenomenologically negligible,
both for W-boson and Z-boson production.

Of course, one could have speculated on this suppression, since the impact
of non-factorizable photonic corrections is already at the level of some $0.1\%$
at NLO. However, the ${\cal O}(\alpha_{\mathrm{s}}\alpha)$ corrections mix EW and QCD
effects, so that small photonic corrections might have been enhanced
by the strong jet recoil effect observed in the $p_{\rT,\Pl}$ distribution.
This enhancement is seen in the virtual and real corrections separately, but not in their sum.
Furthermore, the existence of gluon-induced ($q\Pg$) channels implies
a new feature in the non-factorizable corrections. In the $q\bar q$ channels,
and thus in the full NLO part of the non-factorizable corrections,
the soft-photon exchange proceeds between initial- and final-state particles, whereas
in the $q\Pg$ channels at ${\cal O}(\alpha_{\mathrm{s}}\alpha)$ the photon is also
exchanged between final-state particles. The known suppression mechanisms in 
non-factorizable corrections work somewhat differently in those cases~\cite{Melnikov:1995fx}.

\begin{figure}
\hspace*{1.5em}\epsfig{file=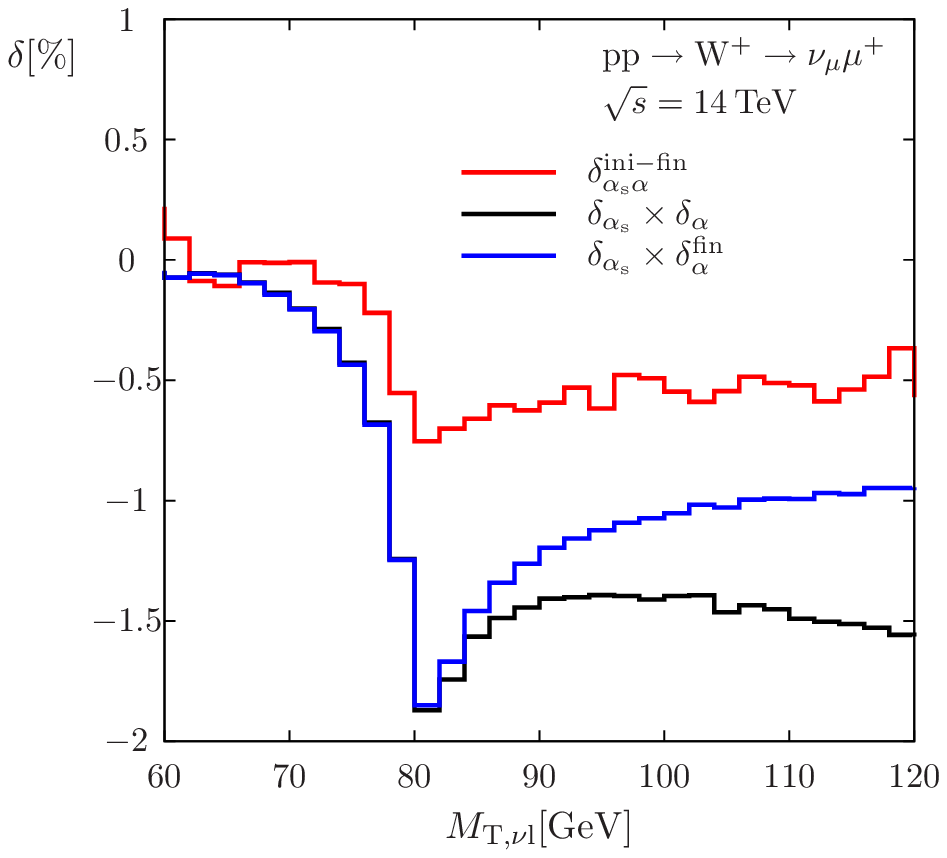,scale=0.7} \hfill 
\epsfig{file=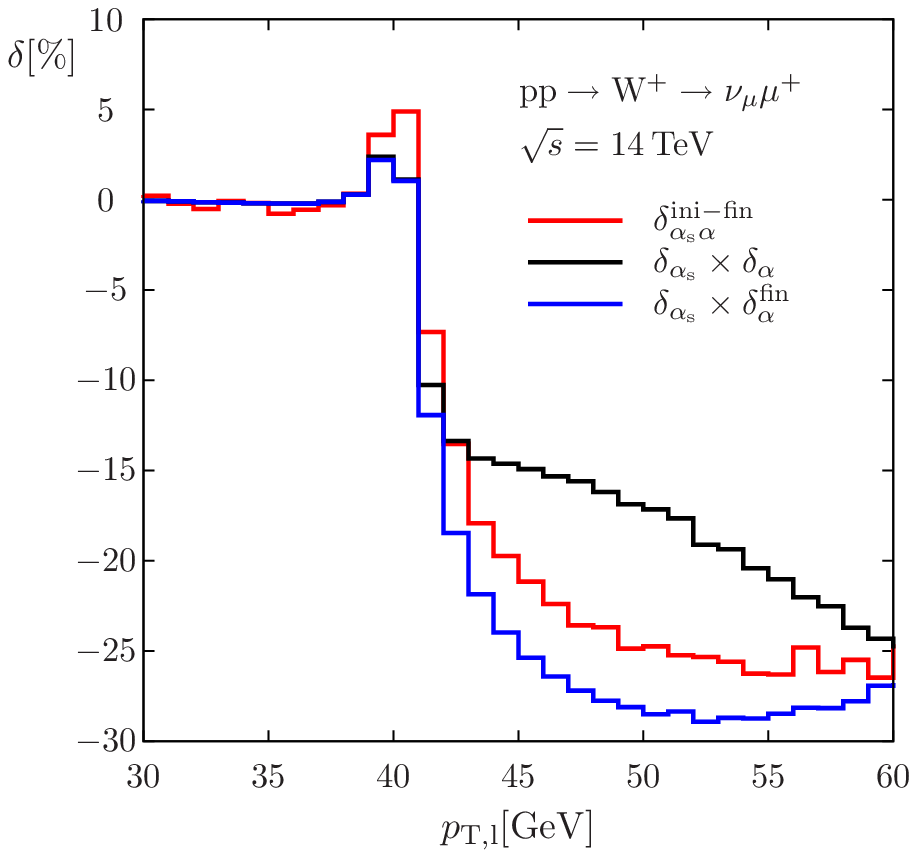,scale=0.7} 
\\[-2.2em]
\caption{Relative factorizable corrections (in red) of ${\cal O}(\alpha_{\mathrm{s}}\alpha)$ 
induced by initial-state QCD and final-state EW contributions 
to the distributions in $M_{\mathrm{T},\nu l}$ (left) and $p_{\mathrm{T,l}}$ (right)
for $\PWp$ production at the LHC. The naive products of the NLO correction
factors $\delta_{\alpha_{\mathrm{s}}}$ and $\delta_{\alpha}$
are shown for comparison (see text).}
\label{fig:Oaas-IFfact}
\vspace*{1.5em}
%
\hspace*{1.5em}\epsfig{file=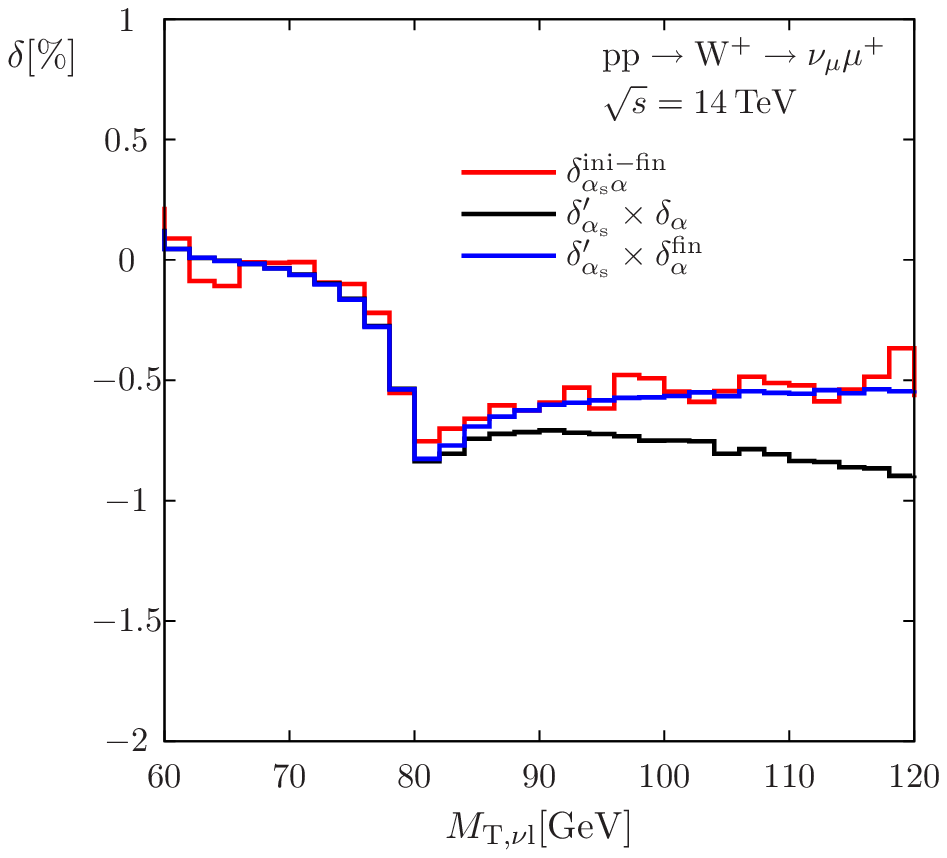,scale=0.7} \hfill 
\epsfig{file=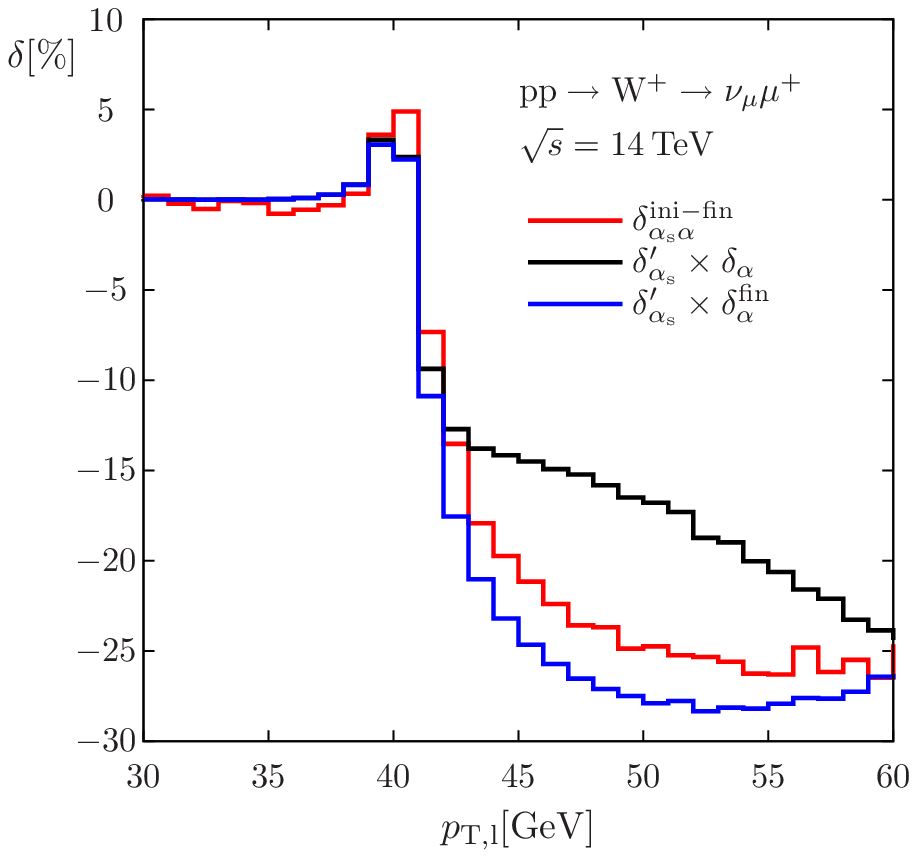,scale=0.7}
\\[-2.2em]
\caption{As in Fig.~5, but with the naive product of QCD and EW 
corrections based on $\delta^{\prime}_{\alpha_{\mathrm{s}}}$ instead 
\looseness -1
of~$\delta_{\alpha_{\mathrm{s}}}$.
}
\label{fig:Oaas-IFfact-2}
\end{figure}

Figures~\ref{fig:Oaas-IFfact} and \ref{fig:Oaas-IFfact-2}
show first preliminary results on the 
``initial--final'' factorizable ${\cal O}(\alpha_{\mathrm{s}}\alpha)$ corrections
$\delta^{\mathrm{ini-fin}}_{\alpha_{\mathrm{s}}\alpha}$
induced by initial-state QCD and final-state EW contributions.
From the results of the PA for the NLO EW corrections, one has to expect that those
contributions furnish the by far dominant part at ${\cal O}(\alpha_{\mathrm{s}}\alpha)$,
while the two other types of factorizable contributions of ``initial--initial'' and 
``final--final'' type are much smaller.
In detail, the figures compare the factorizable initial--final
corrections (red curves) to different versions of 
naive products of the NLO QCD and NLO EW correction factors.
To define the naive products, we write the NLO QCD cross section
$\sigma^{\mathrm{NLO}_{\mathrm{s}}}$ as
\begin{equation}
\sigma^{\mathrm{NLO}_{\mathrm{s}}}
\equiv \sigma^{\mathrm{LO}}(1+\delta_{\alpha_{\mathrm{s}}})
= \sigma^0+\sigma^{\mathrm{LO}}\left(\frac{\sigma^{\mathrm{LO}}-\sigma^0}{\sigma^{\mathrm{LO}}}+\delta_{\alpha_{\mathrm{s}}}\right)
\equiv\sigma^0+\sigma^{\mathrm{LO}}\delta'_{\alpha_{\mathrm{s}}},
\end{equation}
where $\sigma^{\mathrm{LO}}$ and $\sigma^0$ denote the LO cross section
evaluated with LO and NLO PDFs, respectively. The standard QCD $K$~factor is, thus,
given by $K^{\mathrm{NLO}_{\mathrm{s}}}=1+\delta_{\alpha_{\mathrm{s}}}$, and
$\delta'_{\alpha_{\mathrm{s}}}$ differs from $\delta_{\alpha_{\mathrm{s}}}$ by the LO prediction induced by the 
transition from LO to NLO PDFs.
The EW correction factor 
is used in two different versions, first based on the full NLO
correction ($\delta_\alpha$, black curves), and second based on the dominant EW final-state correction
of the PA ($\delta^{\mathrm{fin}}_\alpha$, blue curves). The 
relative EW correction is always derived from the ratio of the NLO EW contribution
$\Delta\sigma^{\mathrm{NLO}_{\mathrm{ew}}}$ and the LO contribution $\sigma^0$ to the NLO
cross section, $\delta_\alpha=\Delta\sigma^{\mathrm{NLO}_{\mathrm{ew}}}/\sigma^0$,
so that the EW correction factors are practically independent of the PDF set.
An ansatz for the cross section $\sigma^{\mathrm{NNLO}_{\mathrm{s\otimes ew}}}_{\mathrm{naive\,fact}}$ 
based on the naive factorization of QCD and EW corrections 
then reads
\begin{equation}
\sigma_{\mathrm{naive\,fact}}^{\mathrm{NNLO}_{\mathrm{s\otimes ew}}}
= \sigma^{\mathrm{NLO}_{\mathrm{s}}}(1+\delta_\alpha)
= \sigma^{\mathrm{LO}}(1+\delta_{\alpha_{\mathrm{s}}})(1+\delta_\alpha).
\end{equation}
This ansatz should approximate the 
full NLO QCD+EW cross section improved by our calculated ${\cal O}(\alpha_{\mathrm{s}}\alpha)$ correction
$\Delta\sigma^{\mathrm{NNLO}_{\mathrm{s\otimes ew}}}_{\mathrm{ini-fin}}$,
\begin{equation}
\sigma^{\mathrm{NNLO}_{\mathrm{s\otimes ew}}}
= \sigma^{\mathrm{NLO}_{\mathrm{s}}}+ \Delta\sigma^{\mathrm{NLO}_{\mathrm{ew}}}
+ \Delta\sigma^{\mathrm{NNLO}_{\mathrm{s\otimes ew}}}_{\mathrm{ini-fin}},
\end{equation}
which is consistently evaluated with NLO PDFs.

Figures~\ref{fig:Oaas-IFfact} and \ref{fig:Oaas-IFfact-2} compare the 
${\cal O}(\alpha_{\mathrm{s}}\alpha)$ correction 
$\delta^{\mathrm{ini-fin}}_{\alpha_{\mathrm{s}}\alpha}
=\Delta\sigma^{\mathrm{NNLO}_{\mathrm{s\otimes ew}}}_{\mathrm{ini-fin}}/\sigma^{\mathrm{LO}}$
with the two versions $\delta_{\alpha_{\mathrm{s}}}\delta_\alpha$
and $\delta^{\prime}_{\alpha_{\mathrm{s}}}\delta_\alpha$, respectively.
The corrections to the W-boson transverse-mass distribution are approximated
by the naive factorization $\delta^{\prime}_{\alpha_{\mathrm{s}}}\delta_\alpha$ quite well. 
The preference of the factorization variant $\delta^{\prime}_{\alpha_{\mathrm{s}}}\delta_\alpha$
over $\delta_{\alpha_{\mathrm{s}}}\delta_\alpha$ can be interpreted upon inspecting the
difference between $\sigma^{\mathrm{NNLO}_{\mathrm{s\otimes ew}}}$ and
the factorized ansatz $\sigma_{\mathrm{naive\,fact}}^{\mathrm{NNLO}_{\mathrm{s\otimes ew}}}$,
\begin{equation}
\frac{\sigma^{\mathrm{NNLO}_{\mathrm{s\otimes ew}}}
-\sigma_{\mathrm{naive\,fact}}^{\mathrm{NNLO}_{\mathrm{s\otimes ew}}}}{\sigma^{\mathrm{LO}}}
= 
\delta^{\mathrm{ini-fin}}_{\alpha_{\mathrm{s}}\alpha}
-\delta'_{\alpha_{\mathrm{s}}}\delta_\alpha,
\end{equation}
i.e.\ the validity of the factorization approximation 
$\sigma_{\mathrm{naive\,fact}}^{\mathrm{NNLO}_{\mathrm{s\otimes ew}}}$
is signalled by a small difference between 
$\delta^{\mathrm{ini-fin}}_{\alpha_{\mathrm{s}}\alpha}$ and
$\delta'_{\alpha_{\mathrm{s}}}\delta_\alpha$.
Naive factorization works for the $M_{\mathrm{T},\nu l}$ distribution, which
is rather insensitive to any recoil effect of the W~boson
with respect to additional jet emission.
The factorization ansatz fails, however,
for the distribution in the lepton transverse momentum, 
which is sensitive
to the interplay between QCD and photonic real-emission effects.
It remains to be seen whether the
${\cal O}(\alpha_{\mathrm{s}}\alpha)$ corrections to the $p_{\mathrm{T,l}}$ distribution
can be reproduced by attaching a photonic shower, which is based on the asymptotics of collinear
photon emission, to a QCD-based prediction.
\looseness-1

The results shown in \reffis{fig:Oaas-IFfact} and \ref{fig:Oaas-IFfact-2}
are obtained by photon recombination, which combines collinear photons and leptons to
a quasi-particle as described in \citeres{Dittmaier:2001ay,Dittmaier:2009cr},
while the NLO EW results shown in \reffi{fig:PA-NLO} involve ``bare muons''
without any photon recombination.
Photon recombination restores the level of inclusiveness required by the KLN theorem
to imply a cancellation of the collinear singularity, which enhances the effect of final-state
radiation. Photon recombination typically reduces the size of photonic corrections by a factor of two.

\looseness-1

\section{Conclusions}

Here we have briefly summarized the main results of \citere{Dittmaier:2014qza}, where
we have shown how the ${\cal O}(\alpha_{\mathrm{s}}\alpha)$
corrections to Drell--Yan processes can be approximated by the leading term in an
expansion about the resonance pole. The quality of such an approximation achieved at
NLO strongly supports the expectation that this approach is sufficient for
observables that are dominated by the resonance, which include, e.g., the ones
relevant for precision determinations of the W-boson mass.
The pole approximation classifies corrections into factorizable and
non-factorizable contributions. Our results show that the latter, which
link production and decay subprocesses via soft-photon exchange, are phenomenologically
negligible. The phenomenologically relevant corrections, thus, are of
factorizable nature and can be attributed to corrections to the initial or final states. 
Again, the pattern of contributions to the pole approximation at NLO gives
a clear picture on the expected hierarchy in higher orders.
At ${\cal O}(\alpha_{\mathrm{s}}\alpha)$, the bulk of corrections will be
contained in the combination of QCD corrections to the production and EW
corrections to the decay processes, with a particular enhancement induced by
final-state radiation off the charged leptons.
Here we have presented first preliminary results on those contributions,
revealing that the naive factorization of NLO QCD and NLO EW corrections approximates
the ${\cal O}(\alpha_{\mathrm{s}}\alpha)$ corrections to the W-boson transverse-mass distribution,
but not the lepton transverse-momentum distribution, which is particularly sensitive
to recoil effects of the W~boson.
The completion of our calculation of ${\cal O}(\alpha_{\mathrm{s}}\alpha)$ 
corrections and the discussion of their phenomenological implications are in progress.

\end{document}

%% file: setup.tex
\newcommand{\lsim}
{\mathrel{\raisebox{-.3em}{$\stackrel{\displaystyle <}{\sim}$}}}

\def\asymp#1%
{\mathrel{\raisebox{-.4em}{$\widetilde{\scriptstyle #1}$}}}

\def\Nequal#1%
{\mathrel{\raisebox{-.5em}{$\stackrel{=}{\scriptstyle\rm#1}$}}}
\newcommand{\dsl}[1]{\not \hspace{-0.7mm}#1}
\def\dsl{\mathpalette\make@slash}
\def\make@slash#1#2{\setbox\z@\hbox{$#1#2$}%
  \hbox to 0pt{\hss$#1/$\hss\kern-\wd0}\box0}

\def\beq{\begin{equation}}
\def\eeq{\end{equation}}
\def\bit{\begin{itemize}}
\def\eit{\end{itemize}}
\def\beqar{\begin{eqnarray}}
\def\eeqar{\end{eqnarray}}
\def\barr#1{\begin{array}{#1}}
\def\earr{\end{array}}
\def\bfi{\begin{figure}}
\def\efi{\end{figure}}
\def\btab{\begin{table}}
\def\etab{\end{table}}
\def\bce{\begin{center}}
\def\ece{\end{center}}

\def\text{\textstyle}

\def\Ga{\Gamma}

\def\Si{\Sigma}

\def\refeq#1{\mbox{(\ref{#1})}}

\def\reffi#1{\mbox{Fig.~\ref{#1}}}
\def\reffis#1{\mbox{Figs.~\ref{#1}}}

\def\citere#1{\mbox{Ref.~\cite{#1}}}
\def\citeres#1{\mbox{Refs.~\cite{#1}}}

\newcommand{\MeV}{\unskip\,\mathrm{MeV}}

\newcommand{\ri}{{\mathrm{i}}}

\newcommand{\rT}{{\mathrm{T}}}

\newcommand{\M}{{\cal{M}}}

\def\mathswitchr#1{\relax\ifmmode{\mathrm{#1}}\else$\mathrm{#1}$\fi}

\newcommand{\PV}{V}
\newcommand{\PW}{\mathswitchr W}

\newcommand{\PZ}{\mathswitchr Z}

\newcommand{\Pg}{\mathswitchr g}

\newcommand{\PWp}{\mathswitchr {W^+}}

\newcommand{\Pl}{l}

\def\mathswitch#1{\relax\ifmmode#1\else$#1$\fi}

\newcommand{\MV}{\mathswitch {M_\PV}}
\newcommand{\MW}{\mathswitch {M_\PW}}

\def\solid{\raise.9mm\hbox{\protect\rule{1.1cm}{.2mm}}}
\def\dash{\raise.9mm\hbox{\protect\rule{2mm}{.2mm}}\hspace*{1mm}}

